\documentclass[3p, 11pt]{elsarticle}
\usepackage{graphicx,amscd,amsmath,amssymb,verbatim}
\usepackage{amsfonts,epsfig}
\usepackage{mathptmx}
\usepackage{setspace}
\usepackage{epsfig}
\usepackage{url}
\usepackage{multirow}
\usepackage{subfigure}
\usepackage{color}

\begin{document}

\journal{Elsevier}

\begin{frontmatter}

\title{A Reactive-Darwinian Model for the Ultimatum Game: On the Dominance of Moderation in High Diffusion}

\author{Roberto da Silva $^{1}$,  Pablo Valverde$^{2}$, Luis C. Lamb$^{3}$}

\address{1-Institute of Physics, Federal University of Rio Grande do Sul,
Av. Bento Gon\c{c}alves, 9500, Porto Alegre, 91501-970, RS, Brazil
{\normalsize{E-mail:rdasilva@if.ufrgs.br, }}}

\address{2 - Escuela de Ciencias F\'{\i}sicas y Matem\'atica, Pontificia Universidad Cat\'olica del Ecuador,
Av.12 de Octubre 1076 y Roca, Quito-Ecuador
{\normalsize{E-mail:pjvalverde@puce.edu.ec}}}

\address{3-Institute of Informatics, Federal University of Rio Grande do Sul,
Av. Bento Gon\c{c}alves, 9500, Porto Alegre, 91501-970, RS, Brazil
{\normalsize{E-mail:lamb@inf.ufrgs.br}}}


\begin{abstract}

We consider a version of the ultimatum game which simultaneously combines 
reactive and Darwinian aspects with offers in [0,1]. By reactive aspects, 
we consider the effects that lead the player to  change their offer given the 
previous result. On the other hand, Darwinian aspects correspond to copying a better 
strategy according to best game payoff when the current player compares with one of 
their neighbours. Therefore, we consider three different strategies, which govern how 
the players change their offers: greedy, moderate, and conservative. First, we provide 
an analytic study of a static version of game, where Darwinian aspects are not 
considered. Then, by using numerical simulations of a detailed and complete 
multi-agent system on a two dimensional lattice, we add an extra feature, in which 
players probabilistically scape from extreme offers (those close to 0 or 1) for 
obvious reasons. The players are also endowed reciprocity on their gains as proposers, 
which is reflected on their gains as responders. We also analyse the influence of the 
player's mobility effects. An analysis of the emergence of coexistence of strategies 
and changes on the dominant strategies are observed, which in turn depends on the 
player's mobility rate.  

\end{abstract}

\end{frontmatter}


\setlength{\baselineskip}{0.7cm}

\section{Introduction}

\label{section:introduction}

How do people agree on a fair salary? How to determine the ideal price of a
product? Or how two members of a species can jointly get some benefits by
combining their efforts? These are clear experimental scenarios of a very
interesting game that mimics bargaining among pairs of different members of
a population, known as the ultimatum game \cite{Guth1982}. The mechanism
underlying this game is very simple: one of the players (the proposer)
proposes a division and a second player (the responder) can either accept or
reject it. If the responder accepts the offer, the values are distributed
according to the division defined by the proposer. Otherwise, both players
earn nothing. The important point here is that the proposer needs the
responder to win (a share of) money, food, objects; the proposer, knowing
that this is a fundamental part for that deal to occur, can impose what part
of the offered quantity he or she wants.

The ultimatum game is a well-known economic experiment. The fairness norm,
sometimes can be inquired among the participants of the game and a costly
punishment (also known as altruistic punishment) can alternatively be framed
as a reduction of payoffs. Further, moral judgements can be important in
some scenarios (see e.g. \cite{Eriksson2017}). So, the fifty-fifty norm
seems to be an important ingredient in the game model, and is completely
contrary to Nash Equilibrium, which establishes that infinitesimal amounts
must be accepted by the responder, but it really does not work in practical
experiments, as offers below one third of the available amount are often
rejected \cite{Szolnoki}. Some authors goes beyond such results, reporting
that minimal offers must be the complementary to the golden ratio, i.e., $%
O=1-\frac{\sqrt{5}-1}{2}\approx \allowbreak 0.381\,97$ \cite{Schuster}, for
offers in the interval $[0,1]$.

Thus, when considering the effects of the ultimatum game on large
populations with or without spatial structures, several perspectives on
social, economic, and biological sense can be explored. Among the
interesting ways to consider this game in large populations, we recently
considered the reactive ultimatum game proposed in \cite{Enock2014,Silva2016}%
. This is an interesting protocol based on direct responses the players can
take given a bad or good result. In this version, we considered that a
player $i$ (the proposer) that puts forward an offer $O_{t}^{(i)}\in \lbrack
0,1]$ at time $t$, which can be accepted or rejected by the other player $j\ 
$(i.e. the responder). The acceptance occurs with probability $%
p_{t}(j|i)=O_{t}^{(i)}$, i.e., the acceptance occurs with higher probability
when the offer is more generous. So the offers can incremented/decremented
by a quantity $\varepsilon >0$ if the responders reject/accept, as a kind of
regulation. In \cite{Silva2016} previous studies of a static version of the
game (fixed strategies) were studied, considering that a player regulates
their offer by following more conservative actions (in the financial sense)
or being more greedy.

However other questions should be answered and explored under this approach,
but first we suggest some reformulation and simplification of the model
considered in \cite{Silva2016}, by reducing from 4 to 3 strategies \footnote{%
What we called a policy in \cite{Silva2016}, here we are naming as a
strategy.} The strategies proposed are: greedy (G), moderate (M), and
conservative (C). Starting from this point, in this work we intend to study
the ultimatum game considering two evolutionary aspects:

\begin{enumerate}
\item \textbf{Evolution of the offers }$O_{t}$, i.e., an evolution based on
reaction, where the player increments/decrements their offer;

\item \textbf{Evolution of the strategies}, where the player can change
their strategy, considering the standard evolutionary game theory based on
Darwinian theory principles.
\end{enumerate}

So our plan in this paper is to perform two types os study. First, we forget
the spatial, Darwinian model, and study a version of iterated reactive
ultimatum game, based only on evolutionary aspect 1, as above. In this case,
the players can mix the three strategies to take their actions, which we
denote as the \emph{static analysis of the game}. We analytically studied
the existence of optimal combination of probabilities of players acting $G$, 
$M$, and $C$, respectively represented by $p_{G}$, $p_{M}$, and $p_{C}$
(with $p_{G}+p_{M}+p_{C}=1$) that results in fair stationary offers.

Second, we consider the model in more complex scenarios by proposing a
Multi-agent System of the Spatial Ultimatum Game (MSSUG). In order to do so,
we consider several possible ingredients for this game: Mobile agents on the
two dimensional lattice that regulate their offers (reactive aspect), and
copy strategies according to the best payoff along the evolution (Darwinian
aspect). Moreover, in this second approach we also consider two additional
ingredients: (a) The players try to avoid both extremely high and extremely
low offers, since the probability of both complete altruism or greedy
characteristics are not interesting in the ultimatum game (due to risk
aversion) and (b) The player not only will take into account the offer when
considering its acceptance, but he/she also compares this offer with their
gains as a proposer (the expected wealth).

We will then show that diffusion of agents in MSSUG leads to dominance of
moderate players showing that extreme actions (be they conservative, or
greedy) do not bring satisfactory stability under Darwinian aspects.

The remainder of the paper is organized as follows. First, we present the
game model. We then separate the model that is considered in the mean field
version from the analytical (static version) and the full (dynamic) model
which includes more features, the MSSUG, to be numerically solved via Monte
Carlo simulations. Our results are divided in two parts. In section \ref%
{Section:ResultsI} we present the results relative to the static aspects of
the model. We present a more refined description than the one preliminarily
presented in \cite{Silva2016}. Numerical solutions of differential equations
are thus obtained. Section \ref{Section:Numerical_simulations} described the
Monte Carlo simulations of the our MSSUG model. Finally, we conclude and
point out open questions in Section \ref{Section:Conclusions}.

\section{The Proposed Model}

\label{Sec:Model}

In this paper the reactive ultimatum game proposed in \cite%
{Enock2014,Silva2016} is studied considering that a player $i\ $(proposer),
puts forward an offer $O_{t}^{(i)}\in \lbrack 0,1]$ at time $t$ that can be
either accepted or rejected by the other player $j\ $(i.e. the responder).
Here, the term reactive means that rejections after playing with a certain
number of players will change the players's opinion about the proposal under
consideration. This will make the players regulate their offers by either
increasing or decreasing them. This game protocol, that regulate actions
after iterations was already used by ourselves in other contexts. For
instance, we used this protocol to model soccer statistics, where the
potential of teams are incremented/decremented after wins/losses determining
the anomalous diffusion which occurs in the statistics of the soccer
championships \cite{soccer1,soccer2,soccer3}.

Here, in a different way, we consider three different strategies and not
four as in \cite{Silva2016}, since two of them are very similar. The
strategies differ on the number of positive responses of the neighbours in
regulating the offers. Basically, each strategy works with a greedy level.
An initial offer $O_{0}$ is assigned equally to all players. At the $t-$th
time step, each player $i=1,...,N$ in a network -- where $N$ is the number
of nodes (that in our particular case is exactly the number of players) --
offers a value for their $k_{i}$ neighbours. A neighbour $j$ accepts the
offer with probability $p(j|i;t)$ at time $t$.

As we compute the number of players that have accepted the proposal $n_{a}(i)
$ and a number $\varepsilon >0$, we have the possible strategies:

\begin{enumerate}
\item \textit{Greedy }($G$)\textbf{:} One acceptance is enough to reduce the
offer - If $n_{a}(i)\geq 1$, then $O_{t+1}^{(i)}=O_{t}^{(i)}-\varepsilon$,
otherwise $O_{t+1}^{(i)}=O_{t}^{(i)}+\varepsilon$;

\item \textit{Moderate }($M$)\textbf{:} Ensures that more than or equal to
half of the neighbours accept the proposal in order to reduce the offer - If 
$n_{a}(i)\geq k_{i}/2$, then $O_{t+1}^{(i)}=O_{t}^{(i)}-\varepsilon$,
otherwise $O_{t+1}^{(i)}=O_{t}^{(i)}+\varepsilon$;

\item \textit{Conservative }($C$)\textbf{:} All neighbours must accept the
proposal to reduce the offer - If $n_{a}(i)=k_{i}$, then $%
O_{t+1}^{(i)}=O_{t}^{(i)}-\varepsilon$, otherwise $%
O_{t+1}^{(i)}=O_{t}^{(i)}+\varepsilon$;
\end{enumerate}

Here, the term \textit{conservative} must be understood by the policy: \emph{%
if you are not completely sure about the acceptance of your offer in the
neighbourhood, you will increase your offer; otherwise you will decrease it.}

In this paper, we consider the reactive ultimatum game under two different
modalities: with and without spatial structure. In the first approach, we
consider an analysis of the simple static combination of strategies,
according to certain probabilities, studying the stationary offer of the
players. In this case we consider that players $i$ and $j$ are uncorrelated
and $p(j|i;t) $ does not depend on $j$ and only on the offer of the
proposer, i.e., $p(j|i;t)=O_{t}^{(i)}$ as considered in \cite{Silva2016}. So
we imagine an iterated game, where a player inserted in a population
executes $N$ iterations (plays or combats) per time step $t$. In each
iteration, it makes offers to $k$ neighbour players (its coordination).
Since the acceptance in this mean field approach does not depend on
characteristics of the responder, but only on the value of the offer, the
iteration is very simple.

In each iteration the player takes one strategy $G$, $M$, or $C$ with
respective probabilities $p_{G}$, $p_{M}$, and $p_{C}$, such that $%
p_{G}+p_{M}+p_{C}=1$, and the question here is, what value of stationary
offer 
\begin{equation*}
\left\langle O\right\rangle _{\infty }=\lim_{t\rightarrow \infty }\frac{1}{N}%
\sum_{i=1}^{N}O_{t}^{(i)}
\end{equation*}%
is obtained after this iteration? In our considerations, in this mean field
analysis of the model we performed some analytical predictions subdividing
the analysis in a particular case $k=4$ and considering $k$ as arbitrary.
Numerical simulations in this case were performed only by a double check
since we analytically describe this case.

In a second analysis we consider that players occupy a two dimensional
lattice, where each site has a player associated to it (there are no empty
sites). Each player makes a proposal to their four nearest neighbours and
they will be responder of their four nearest neighbours when these players
act as proposers -- which establishes a symmetric game (in this case we
consider only $k=4$, for the sake of simplicity, but extensions should be
analysed). In order to explore more challenges, we consider some innovations
which define the MSSUG:

\begin{enumerate}
\item \textbf{MSSUG ingredient}: \textit{Accepting depends on both players}:
The responder desires to win a quantity similar to the one obtained when
they are a proposer. Thus, we introduce a correlation between the acceptance
according to this point of view and now the acceptance depends not only on
the offer magnitude, but also on the expected wealth of the responder: 
\begin{equation*}
p(j|i)=O_{t}^{(i)}f_{ij}(t)
\end{equation*}%
where 
\begin{equation*}
f_{ij}(t)=\left\{ 
\begin{array}{cl}
1 & \text{if\ }O_{t}^{(j)}\geq 1-O_{t}^{(i)} \\ 
&  \\ 
0 & \text{otherwise}%
\end{array}%
\right.
\end{equation*}

\item \textbf{MSSUG ingredient: }\textit{Scape from extreme situations}: The
players scape from extremes which is a kind of regulatory mechanism to avoid
gaining very low quantities in extreme altruism or simply never gaining due
to extreme greedy. In this case, changes in the reduction or increase of the
offer must respect boundaries. The player $i$ will decrease the offer with
probability $O_{t}^{(i)}$ and will increase the offer with probability $%
1-O_{t}^{(i)}$, regulating the strategies previously defined. The idea here
is very simple: small offers and large offers, respectively, decrease the
probability of decrease and increase of the offers, escaping from extremes;

\item \textbf{MSSUG ingredient: }\textit{Darwinian Copy}:\textbf{\ }Each
player is born with strategy randomly chosen with same probability (1/3).
After the \textquotedblleft combats\textquotedblright\ (i.e., every player
has offered to its nearest four neighbour and has responded to the same four
neighbours), every player draws a nearest neighbour and copy her strategy if
her payoff is greater than their own payoff (Darwinian copy);

\item \textbf{MSSUG ingredient: }\textit{Mobile agents}:\textbf{\ }After all
players have made their offers to their respective neighbours and also have
responded to their proposers, and have performed the copy (or not) of a new
strategy, we draw $N$ different neighbour pairs on the lattice, and we swap
this position with probability $\alpha$, i.e., the players can diffuse.
After all these procedures, we say that one simulation time step is
performed. The players can diffuse on the lattice and swap positions since
the lattice is complete.
\end{enumerate}

If the first analysis (static) we bring interesting points about
optimization of the portfolio of static strategies. On the other hand, with
these new ingredients, considering agents on the lattice (MSSUG), other
details can be studied considering the coexistence, survival, and
fluctuations of strategies under different values of the mobility parameters 
$\alpha $. The most important question here is related to the significance
of $\alpha $ in changing the survival of the three strategies previously
defined. We will observe that as $\alpha $ enlarges, the moderate agents
dominate the other strategies after a period of coexistence with the
conservative strategy. The effects of $\alpha $ also are analysed on the
average payoff, average offers, and payoff distribution measured by the Gini
coefficient of the population of agents.

\section{Results I: Static Aspects}

\label{Section:ResultsI}

Understanding the reasons which lead to fair behaviour in the iterated
ultimatum game under a single fixed strategy, or considering players with a
fixed (static) mixing of strategies can be very interesting. In previous
works about the ultimatum game, we performed detailed studies of the game in
large populations, considering probabilistic different rules for offer
proposal and acceptance \cite{SilvaJTB,SilvaBJP}.

Here, we also study the current model by first considering this static
version of the game. IN order to do so, at time $t$ of iteration let us
suppose that the player's offer is $O_{t}$ and, in a first analysis, we are
considering that this player interacts with $k=4$ other players in each
play. Since we are supposing that $N$ plays are performed per time unit, in
a \textquotedblleft mean-field\textquotedblright\ analysis, we consider $%
\left\langle O_{t}\right\rangle $ instead of $O_{t}$ for approximation.
Thus, recurrence relations can be written for the average variation on the
offer increment and, for example, we consider that the average variation on
the offer increment of a moderate proposer per combat can be easily
calculated according to a dynamics purely reactive:%
\begin{equation}
\Delta \left\langle O_{t}\right\rangle _{M}=p(n_{a}=\left\{ 2,3,4\right\}
)(-\varepsilon )+p(n_{a}=\left\{ 0,1\right\} )(+\varepsilon )
\end{equation}%
sure $p(n_{a}=\{2,3,4\})+p(n_{a}=\{0,1\})=1$, such that 
\begin{equation}
\Delta \left\langle O_{t}\right\rangle _{M}=\left(
2p(n_{a}=\{0,1\})-1\right) \varepsilon
\end{equation}

But%
\begin{equation}
\begin{array}{lll}
p(n_{a}=\{0,1\}) & = & \sum_{n_{a}=0}^{1}\frac{4!\left\langle
O_{t}\right\rangle ^{n_{a}}(1-\left\langle O_{t}\right\rangle )^{4-n_{a}}}{%
n_{a}!(4-n_{a})!} \\ 
&  &  \\ 
& = & -3\left\langle O_{t}\right\rangle ^{4}+8\left\langle
O_{t}\right\rangle ^{3}-6\left\langle O_{t}\right\rangle ^{2}+1\allowbreak%
\end{array}%
\end{equation}

Resulting in 
\begin{equation}
\Delta \left\langle O_{t}\right\rangle _{M}=\left( -6\left\langle
O_{t}\right\rangle ^{4}+16\left\langle O_{t}\right\rangle
^{3}-12\left\langle O_{t}\right\rangle ^{2}+1\right) \varepsilon
\end{equation}

Similarly we also conclude that for a greedy player:

\begin{equation}
\begin{array}{lll}
\Delta \left\langle O_{t}\right\rangle _{G} & = & \left(
2p(n_{a}=0)-1\right) \varepsilon \\ 
&  &  \\ 
& = & \left[ 2(1-\left\langle O_{t}\right\rangle )^{4}-1\right] \varepsilon 
\text{,} \\ 
&  &  \\ 
& = & \left[ 2\left\langle O_{t}\right\rangle ^{4}-8\left\langle
O_{t}\right\rangle ^{3}+12\left\langle O_{t}\right\rangle ^{2}-8\left\langle
O_{t}\right\rangle +1\right] \varepsilon \text{,}%
\end{array}%
\end{equation}%
and for a conservative one:

\begin{equation}
\begin{array}{lll}
\Delta \left\langle O_{t}\right\rangle _{C} & = & \left(
1-2p(n_{a}=4)\right) \varepsilon \\ 
&  &  \\ 
& = & (1-2\left\langle O_{t}\right\rangle ^{4})\varepsilon \text{,}%
\end{array}%
\end{equation}

If we consider that $p_{G}$, $p_{M}$, and $p_{C}$ are the probabilities of a
player chosen at random acting as a greedy, moderate, conservative player
respectively, such that $p_{G}+p_{M}+p_{C}=1$. So we can write a recurrence
relation to the offer the players considering a population where the players
act according o these probabilities:%
\begin{equation*}
\left\langle O_{t+1}\right\rangle \approx \left\langle O_{t}\right\rangle
+\rho _{G}\Delta \left\langle O_{t}\right\rangle _{G}+\rho _{M}\Delta
\left\langle O_{t}\right\rangle _{M}+\ \rho _{C}\Delta \left\langle
O_{t}\right\rangle _{C}
\end{equation*}

Expanding and grouping we have: 
\begin{equation}
\begin{array}{lll}
\left\langle O_{t+1}\right\rangle -\left\langle O_{t}\right\rangle & = & 
\left( 2\rho _{G}-6\rho _{M}-2\rho _{C}\right) \varepsilon \left\langle
O_{t}\right\rangle ^{4} \\ 
&  & +(16\rho _{M}-8\rho _{G}\allowbreak )\varepsilon \left\langle
O_{t}\right\rangle ^{3}+12(\rho _{G}-\rho _{M})\varepsilon \left\langle
O_{t}\right\rangle ^{2} \\ 
&  & -8\rho _{G}\varepsilon \left\langle O_{t}\right\rangle +\varepsilon 
\text{. }%
\end{array}%
\end{equation}

So considering that $\lim_{\tau \rightarrow 0}\frac{\left\langle
O_{t+1}\right\rangle -\left\langle O_{t}\right\rangle }{\tau }=\frac{%
d\left\langle O_{t}\right\rangle }{dt}$, and so that the changes of offers
occur such that $\tau =\varepsilon $, we have the equation that governs the
evolution of offer 
\begin{equation}
\frac{d\left\langle O_{t}\right\rangle }{dt}=A\left\langle
O_{t}\right\rangle ^{4}+B\left\langle O_{t}\right\rangle ^{3}+C\left\langle
O_{t}\right\rangle ^{2}+D\left\langle O_{t}\right\rangle +1
\label{Eq:evolucionary_differential_equation_k=4}
\end{equation}%
where%
\begin{equation*}
\begin{array}{lll}
A & = & 2(p_{G}-p_{C})-6\rho _{M} \\ 
&  &  \\ 
B & = & 16p_{M}-8p_{G} \\ 
&  &  \\ 
C & = & 12(p_{G}-p_{M}) \\ 
&  &  \\ 
D & = & -8p_{G}%
\end{array}%
\end{equation*}

The solutions of the equation \ref{Eq:evolucionary_differential_equation_k=4}
can be explored to perform comparisons with Numerical Monte Carlo
simulations in two dimensional lattices, which will further be performed
when we explore the MSSUG, which includes all ingredients of diffusion with
swaps of the agents, and also Darwinian aspects, instead of purely reactive
evolutionary aspects considered in the mean-field equations.

In the next section, we will explore some analytical solutions of the Eq. %
\ref{Eq:evolucionary_differential_equation_k=4} and its extension for
arbitrary $k$, according to some conditions (portfolios).

\subsection{Case studies}

In this subsection we will explore some analytical results first for $k=4$
and after for arbitrary $k$.

\subsubsection{$k=4$}

Let us starting our study, where the players always act with the same police:

\textbf{Situation I}: \textit{Acting} \textit{only a conservative player}: $%
p_{C}=1$, and $p_{M}=p_{G}=0$

Thus from Eq. \ref{Eq:evolucionary_differential_equation_k=4}, we have 
\begin{equation}
\frac{d\left\langle O_{t}\right\rangle }{dt}=1-2\left\langle
O_{t}\right\rangle ^{4}  \label{Eq:onlyhc}
\end{equation}%
and so by integrating this differential equation, we have that its solution
is given by: 
\begin{equation*}
K(\left\langle O_{t}\right\rangle )=K(O_{0})+t
\end{equation*}%
where $K(z)=2^{\frac{3}{4}}\left( \frac{1}{4}\arctan 2^{1/4}z-\frac{1}{8}\pi
-\frac{1}{8}\ln \frac{\left( z-2^{-1/4}\right) }{\left( z+2^{-1/4}\right) }%
\right) $.

Since $\arctan z=\frac{1}{2}\ln \frac{(1+z)}{(1-z)}$ (principal value), thus 
$\arctan 2^{1/4}z=\frac{1}{2}\ln \frac{(z+2^{-1/4})}{(-z+2^{-1/4})}$, and 
\begin{equation*}
K(z)=2^{\frac{3}{4}}\left[ \frac{1}{4}\ln \frac{z+2^{-1/4}}{z-2^{-1/4}}-%
\frac{1}{8}\pi (1-i)\right]
\end{equation*}

Thus, after some straightforward algebra we have 
\begin{equation*}
\left\langle O_{t}\right\rangle =2^{-1/4}\frac{\left[ \left( \frac{%
O_{0}+2^{-1/4}}{O_{0}-2^{-1/4}}\right) \exp (2^{5/4}t)+1\right] }{\left[
\left( \frac{O_{0}+2^{-1/4}}{O_{0}-2^{-1/4}}\right) \exp (2^{5/4}t)-1\right] 
}
\end{equation*}

We can observe that $\lim_{t\rightarrow \infty }\left\langle
O_{t}\right\rangle =2^{-1/4}$ independently from $O_{0}$. Moreover, this is
exactly the solution of $1-2\left\langle O_{\infty }\right\rangle ^{4}=0$,
i.e., $\left\langle O_{\infty }\right\rangle =2^{-1/4}\approx \allowbreak
0.840\,90$, corresponding to a stationary solution of a differential
equation.

\textbf{Situation II}: \textit{Acting only as a moderate player}: $p_{M}=1$,
and $p_{G}=p_{C}=0$

In this case we have 
\begin{equation}
\frac{d\left\langle O_{t}\right\rangle }{dt}=-6\left\langle
O_{t}\right\rangle ^{4}+16\left\langle O_{t}\right\rangle
^{3}-12\left\langle O_{t}\right\rangle ^{2}+1  \label{Eq:onlyconservatives}
\end{equation}%
which has not exact solution, since we have no knowledge about the primitive 
$\int \frac{dx}{-6x^{4}+16x^{3}-12x^{2}+1}$, but has a stationary solution
under condition: $-6\left\langle O_{t}\right\rangle ^{4}+16\left\langle
O_{t}\right\rangle ^{3}-12\left\langle O_{t}\right\rangle ^{2}+1=0$. It can
be checked that only acceptable critical point is 
\begin{equation}
O_{\infty }\approx 0.385\,73
\end{equation}%
which again suggests a no dependence of $O_{0}$.

\textbf{Situation III}: \textit{Acting only as a greedy player}: $p_{G}=1$,
and $p_{C}=p_{M}=0$

Thus, we have:

\begin{equation}
\frac{d\left\langle O_{t}\right\rangle }{dt}=2\left\langle
O_{t}\right\rangle ^{4}-8\left\langle O_{t}\right\rangle ^{3}+12\left\langle
O_{t}\right\rangle ^{2}-8\left\langle O_{t}\right\rangle +1
\label{Eq:onlygreedy}
\end{equation}

After some algebra, by integrating this equation we have

\begin{equation*}
\left\langle O_{t}\right\rangle =1+2^{-1/4}\frac{1+\left( \frac{%
2^{1/4}(O_{0}-1)-1}{2^{1/4}(O_{0}-1)+1}\right) \exp (2^{5/4}t)}{1-\left( 
\frac{2^{1/4}(O_{0}-1)-1}{2^{1/4}(O_{0}-1)+1}\right) \exp (2^{5/4}t)}
\end{equation*}%
where we can verify that $\lim_{t\rightarrow \infty }\left\langle
O_{t}\right\rangle =1-2^{-1/4}\approx 0.1591$.

Again we can observe that stationary solution does not depend on the initial
condition $O_{0}$, since the only real acceptable ($0\leq O_{\infty }\leq 1$%
) of 
\begin{equation*}
2O^{4}-8O^{3}+12O^{2}-8O+1=0
\end{equation*}%
is $O_{\infty }=1-2^{-\frac{1}{4}}\approx 0.159\,1$, exactly as expected.

\textbf{Situation IV}: \textit{Mixing strategy}: $p_{M}=p_{C}=p_{G}=1/3$

This situation is important. When the players can choose among the three
different polices with same probability, how the stationary offer works? For
these conditions we have

\begin{equation}
\frac{d\left\langle O_{t}\right\rangle }{dt}=-2\left\langle
O_{t}\right\rangle ^{4}+\frac{8}{3}\left\langle O_{t}\right\rangle ^{3}-%
\frac{8}{3}\left\langle O_{t}\right\rangle +1
\end{equation}

The critical points (stationary solution) is obtained by solving: $%
-2\left\langle O_{t}\right\rangle ^{4}+\frac{8}{3}\left\langle
O_{t}\right\rangle ^{3}-\frac{8}{3}\left\langle O_{t}\right\rangle +1=0$.
The only acceptable solution is \ $\left\langle O_{\infty }\right\rangle
=0.428\,34$.

Now an important point must be considered. It is tempting to think that such
result is the simple mean of the results from situation I, II, and III. But
this is not the case since $\overline{O_{\infty }}=(\left\langle O_{\infty
}\right\rangle _{I}+\left\langle O_{\infty }\right\rangle _{II}+\left\langle
O_{\infty }\right\rangle _{III})/3=0.461\,91$. This result corresponds to
the situation where $\rho _{G}=\rho _{C}=\rho _{M}=1/3$ are the densities of
players of each kind in a population and that these players keep their
strategies fixed along time.

\subsubsection{Optimal combination of probabilities for completely fair
stationary offers}

An interesting question is how a player must combine their strategies to
obtain fair offers along time, offering and therefore gaining similar
quantities? To answer to this question, we can observe that the triple $%
(p_{G},p_{C},p_{M})$ that solve: $AO_{\infty }^{4}+BO_{\infty
}^{3}+CO_{\infty }^{2}+DO_{\infty }+1=0$, for $O_{\infty }=1/2$, leads to
equation: 
\begin{equation}
15p_{G}+11p_{M}+p_{C}=8
\end{equation}%
under restriction $p_{G}+p_{M}+p_{C}=1$. Combining these equations we have, 
\begin{equation}
p_{C}=\frac{2}{5}p_{G}+\frac{3}{10},  \label{Eq:optimal}
\end{equation}%
which leads to $0\leq p_{G}\leq 1$. So for any $p_{M}$, we obtain a $p_{G},$%
such that $0\leq p_{G}\leq \frac{1}{2}$, which $\rho _{C}=1-$ $\rho
_{G}-\rho _{M}$ and this triple essentially determined by the relation \ref%
{Eq:optimal} determines a stationary completely fair offer, $O_{\infty }=1/2$%
. For example, for $\rho _{M}=0$, if the player acts half the time as a
conservative and half the time as a greedy player we have the fair offer at
long periods of time.

On the other hand, we can have more exotic choices. For example, if $p_{M}=%
\frac{6}{10}$, we have $p_{G}=\allowbreak \frac{1}{14}$ and $\rho
_{C}=\allowbreak \frac{23}{70}$, which is a hybrid strategy.

\begin{figure}[tbh]
\begin{center}
\includegraphics[width=1.0\columnwidth]{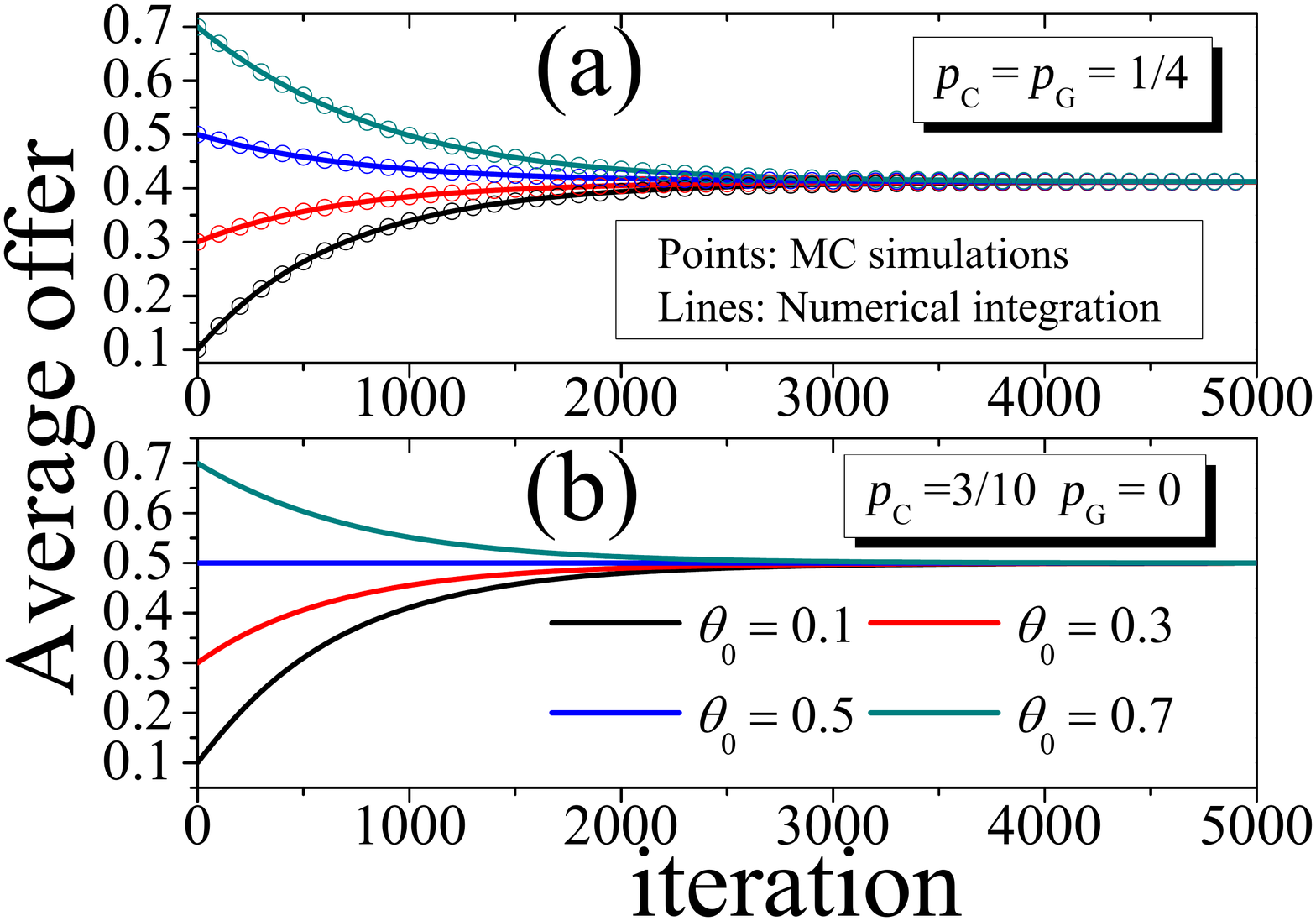}
\end{center}
\caption{Evolution in time of the average offer for two different cases: the
first one, the player adopts the strategies with probabilities $%
p_{C}=p_{G}=1/4$. Just for comparison, we also study the temporal evolution
of the average offer for a case where the steady state is analytically
expected to be fair: $\left\langle O_{\infty }\right\rangle =1/2$, we choose 
$p_{G}=0$ and $p_{C}=3/10$ for the sake of simplicity. In both cases we
considered four different initial offers. In the first case, we can observe
that a stationary fair offer is not obtained unlike the second situation. We
observe that the initial condition is not important to the steady state of
the average offer. The points in the first case show simple MC simulations
performed to check the exact result. }
\label{Fig:static_MCxEDP}
\end{figure}

In order to check such results, we obtain the temporal evolution of the
average offer, by solving the temporal evolution equations, for two
different cases: In the first one Fig. \ref{Fig:static_MCxEDP} (a), the
player adopts the strategies with probabilities $p_{C}=p_{G}=1/4$. Thus, for
a simple comparison, we also study the time evolution of the average offer
for a case where the steady state is analytically expected to be fair: $%
\left\langle O_{\infty }\right\rangle =1/2$, we choose a particular case
from Eq. \ref{Eq:optimal}: $p_{G}=0$ and $p_{C}=3/10$, for the sake of
simplicity Fig. \ref{Fig:static_MCxEDP} (b). In both cases we considered
four different initial offers. In the first case we can observe that the
stationary fair offer is not observed, actually $\left\langle O_{\infty
}\right\rangle \approx 0.4$, which differs on the second situation, where we
check $\left\langle O_{\infty }\right\rangle \approx 0.5$. We observe that
the initial condition is not important to the steady state of the average
offer. The points in the first case correspond to simple MC simulations
performed for checking the exact result.

\begin{figure}[tbh]
\begin{center}
\includegraphics[width=1.0\columnwidth]{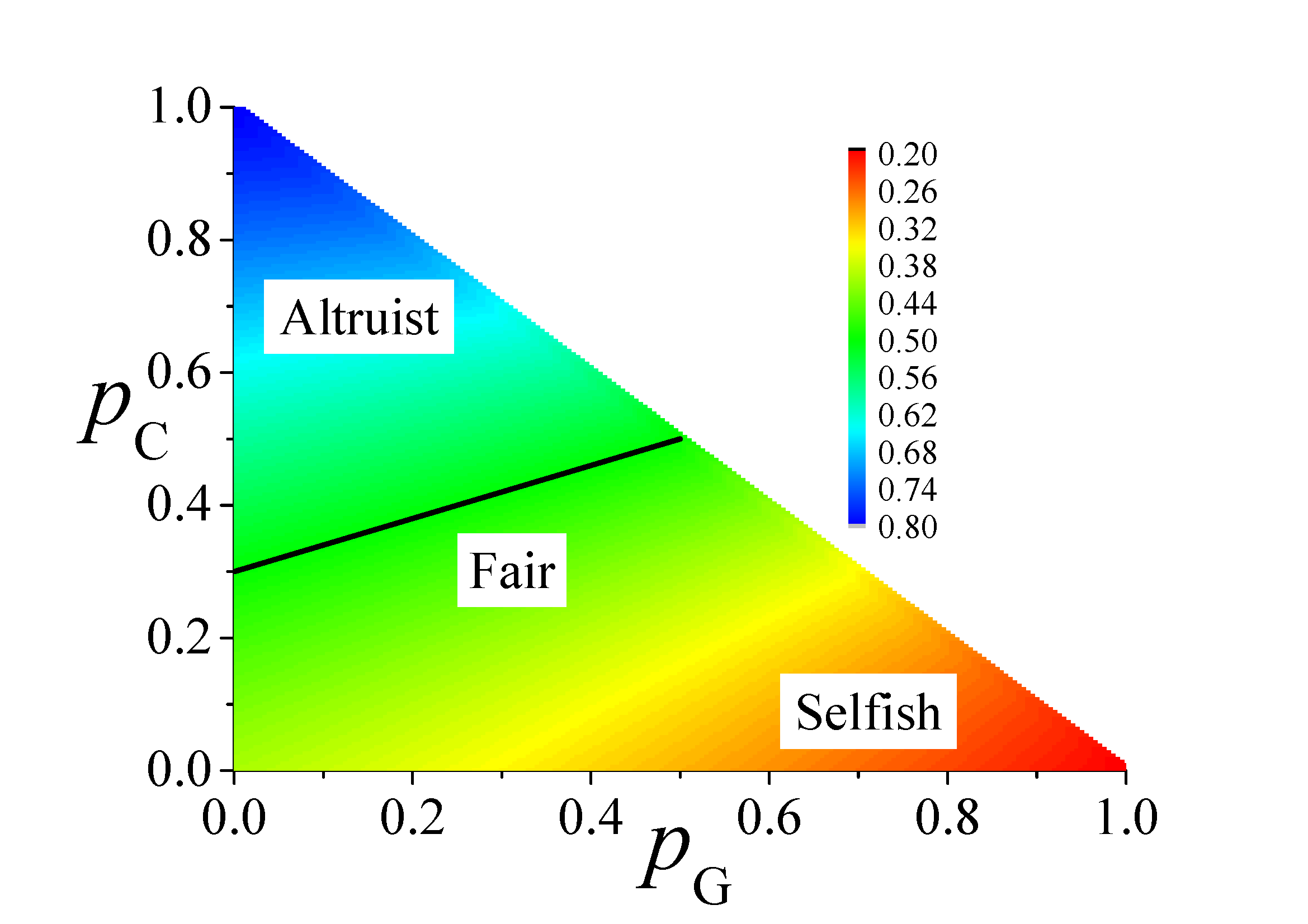}
\end{center}
\caption{Colour map, steady state average offer for all possible pairs $%
(p_{G},p_{C}$). We have 3 distinct regions: altruist, fair, and selfish one.
The dark line corresponds to exact fairness condition given by Eq. \protect
\ref{Eq:optimal}.}
\label{Fig:Color_map}
\end{figure}

Since we studied the evolution of the average offers, and we also verified
the no dependence on initial offer, we can study in further detail the
fairness condition predicted by Eq. \ref{Eq:optimal}. Thus, we solve Eq. \ref%
{Eq:evolucionary_differential_equation_k=4} for all possible pairs $\left(
p_{G},p_{C}\right) $, and the steady state average offer is drawn in Fig. %
\ref{Fig:Color_map}.

We can observe a complete description of the steady state average offer
divided in three regions $\left\langle O_{\infty }\right\rangle <1/2$
(selfish region), $\left\langle O_{\infty }\right\rangle >1/2$ (altruist
region), and finally $\left\langle O_{\infty }\right\rangle \approx 1/2$
(fair region) which contains the optimal fairness line given by Eq. \ref%
{Eq:optimal}.

\subsection{Arbitrary $k$ and preliminary numerical investigations}

It is important to observe that we can generalize Eq. \ref%
{Eq:evolucionary_differential_equation_k=4} for an arbitrary coordination $k$%
. By following similar steps we can deduce that%
\begin{equation}
\frac{d\left\langle O_{t}\right\rangle }{dt}=\left( 2\sum_{m=0}^{k/2-1}\frac{%
k!\left\langle O_{t}\right\rangle ^{m}(1-\left\langle O_{t}\right\rangle
)^{k-m}}{m!(k-m)!}-1\right) p_{M}+\left( 2(1-\left\langle O_{t}\right\rangle
)^{k}-1\right) p_{G}+\left( 1-2\left\langle O_{t}\right\rangle ^{k}\right)
p_{C}
\end{equation}

Under stationary condition $\lim_{t\rightarrow \infty }\left\langle
O_{t}\right\rangle =O_{\infty }=1/2$, we have for large $k$:%
\begin{equation}
\begin{array}{lll}
\sum_{m=0}^{k/2-1}\frac{k!\left\langle O_{t}\right\rangle
^{m}(1-\left\langle O_{t}\right\rangle )^{k-m}}{m!(k-m)!} & = & \frac{1}{%
2^{k}}\sum_{m=0}^{k/2-1}\binom{k}{m} \\ 
&  &  \\ 
& \approx & \frac{2}{\sqrt{2\pi k}}\int_{0}^{k/2}dm\ e^{-\frac{(m-k/2)^{2}}{%
k/2}} \\ 
&  &  \\ 
& = & 1/2%
\end{array}%
\end{equation}%
and since $d\left\langle O_{t}\right\rangle /dt=0$, we have: 
\begin{equation}
\frac{p_{G}}{p_{C}}=1
\end{equation}

\begin{figure}[tbh]
\begin{center}
\includegraphics[width=1.0\columnwidth]{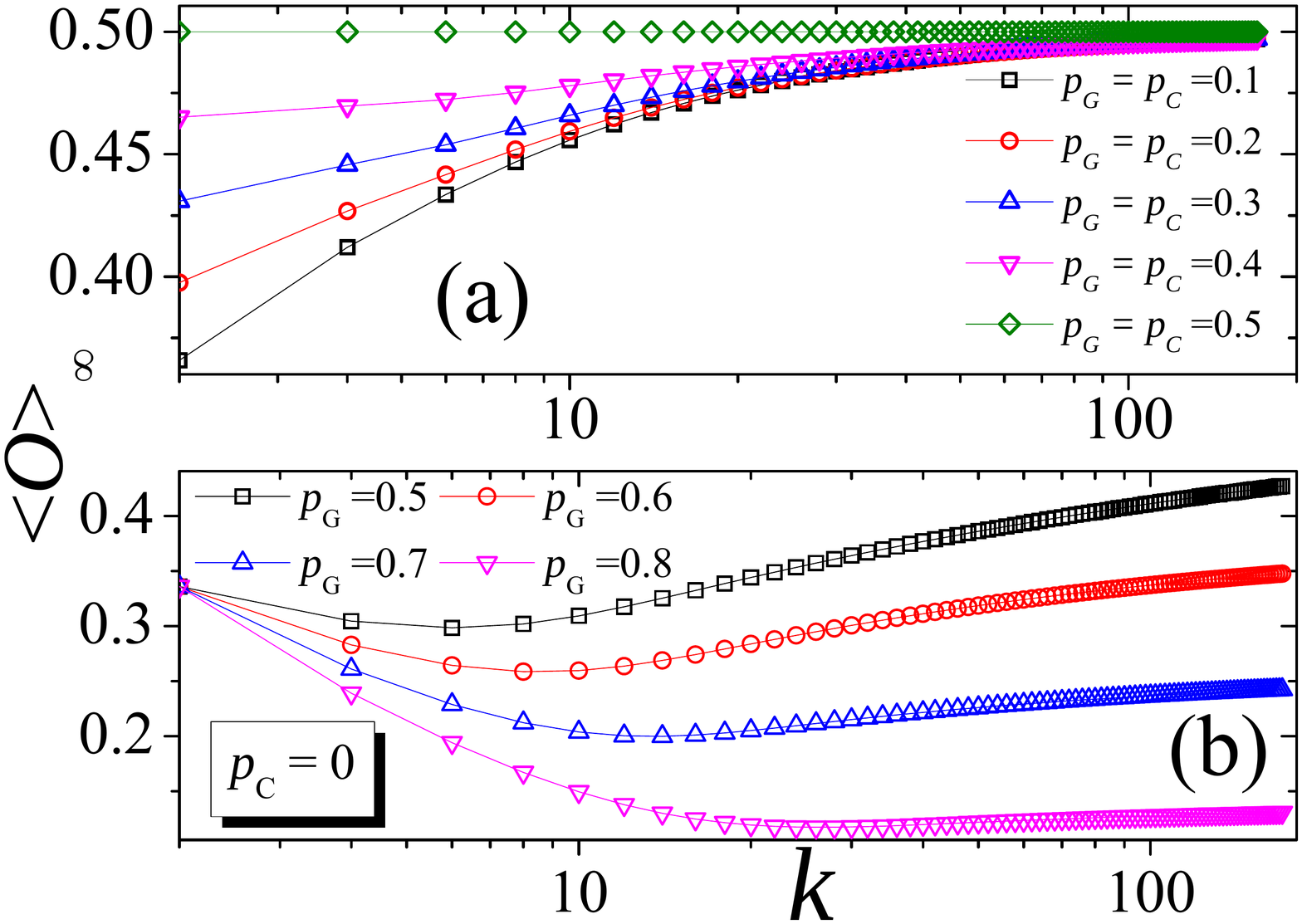}
\end{center}
\caption{Steady state average offer as function of $k$. (a) Fairness occurs
for high $k$ given a player that has equal probabilities to act as
conservative and greedy, unlike in their moderate actions. (b) As the number
of greedy players increases, the steady state average offer decreases, but
there is a $k_{c}$ for that one and the steady state average offer is
minimal.}
\label{Fig:average_offer_vs_k}
\end{figure}

This result is very important but also intuitive. For any probability of a
player acting as a moderate player, if you consider equal probabilities of
acting as greedy and conservative, the player always leads to a fair offer
when she/he plays with a large sample of players taking their decisions.
This is expected, since as $k$ grows, the probability of a greedy player
decreasing their offer is high, while the probability that a conservative
player increases their offer is also high, if a player acts with equal
probability for greedy and conservative, they must keep a fair steady state
average offer, unlike in their moderate actions. This can be observed in
Fig. \ref{Fig:average_offer_vs_k} (a).

This figure depicts the steady state average offer for different situations
of $p_{G}=p_{C}$. However, for high $k$, the probability of the offer
decreasing must be very small, and therefore we consider a particular, but
interesting case $p_{C}=0$. So we study $\left\langle O_{\infty
}\right\rangle $ as function of $k$ and the results is surprising. Although, 
$\left\langle O_{\infty }\right\rangle $ decreases as $p_{G}$ increases, for
fixed $p_{G}$($p_{M}$), there is a $k_{C}$ such that $\left\langle O_{\infty
}\right\rangle $ is minimal as shown in Fig. \ref{Fig:average_offer_vs_k}
(b). This non-monotonic behaviour in the static version can be explained
taking into account a complex scenario. There is a competition of the
parameters for the greedy actions. For intermediate $k$ and for good offers,
a sequence of decreases on the offer can occur. At large $k$, it should be
easy to find an accepting, but the previous catastrophic sequence of
decreases makes the offers not so attractive and this balance between large $%
k$ and low offers must be the reason of the non-monotonic behaviour. On the
other hand, moderate actions are not sensitive to $k$.

Since we have studied some interesting aspects of the static ultimatum game,
in the next section we present the results of our MSSUG in the two
dimensional lattice, exploring the points related to survival of strategies
with introduction of mobility of the players.

\section{Results II: Numerical simulations}

\label{Section:Numerical_simulations}

We carried out numerical simulations of the MSSUG in two-dimensional
lattices ($k=4$). Each player starts with one of three strategies, which are
equally probable. Thus each player interacts with their four neighbours,
acting as proposer. The responder acts analogously with respect to their
four neighbours. The update is synchronous, i.e., after all interactions the
strategies are updated for all players at once.

\begin{figure}[tbh]
\begin{center}
\includegraphics[width=1.0%
\columnwidth]{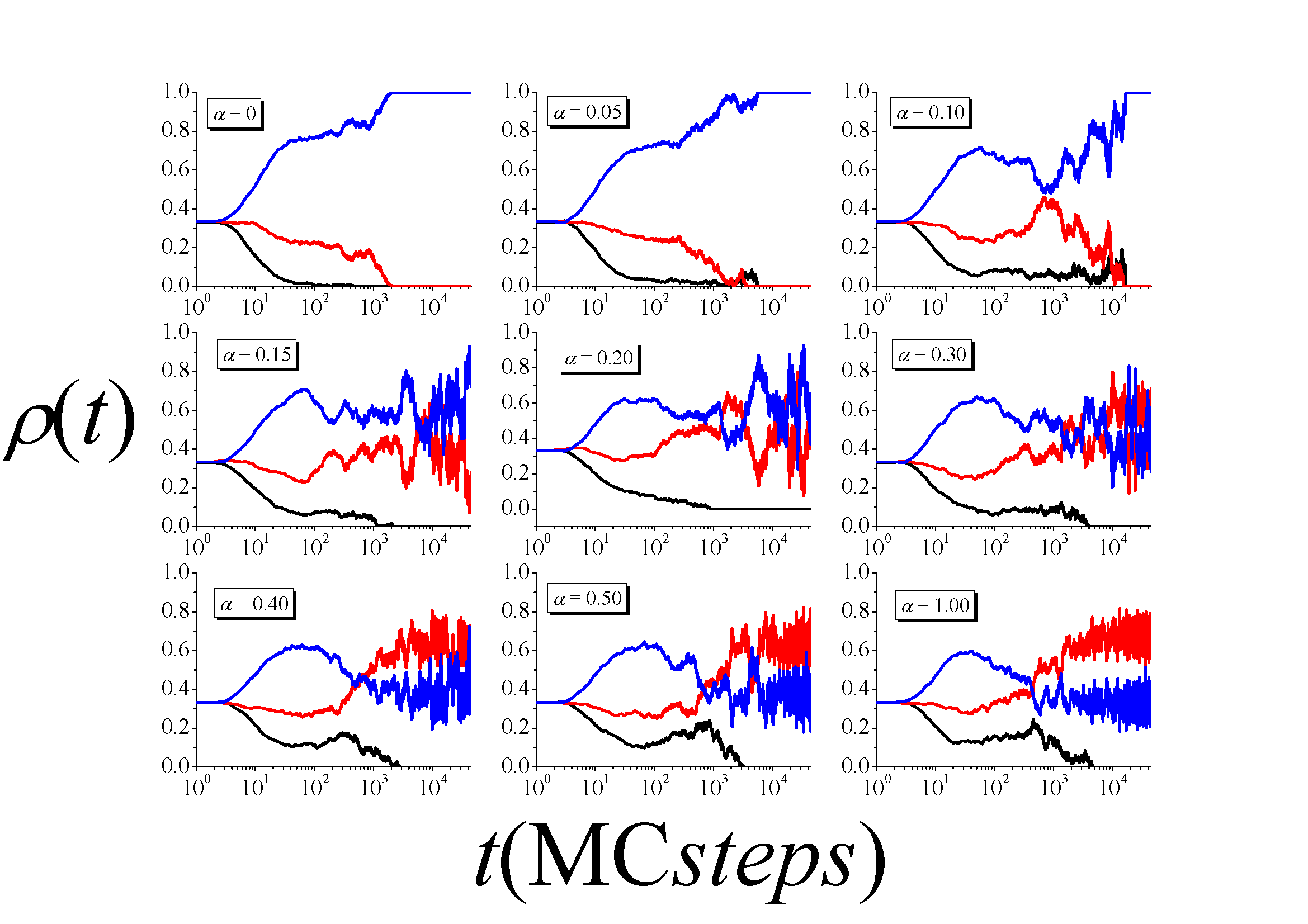}
\end{center}
\caption{Density of strategies of the MSSUG for one only run. }
\label{Fig:one_run}
\end{figure}

\begin{figure}[tbh]
\begin{center}
\includegraphics[width=1.0\columnwidth]{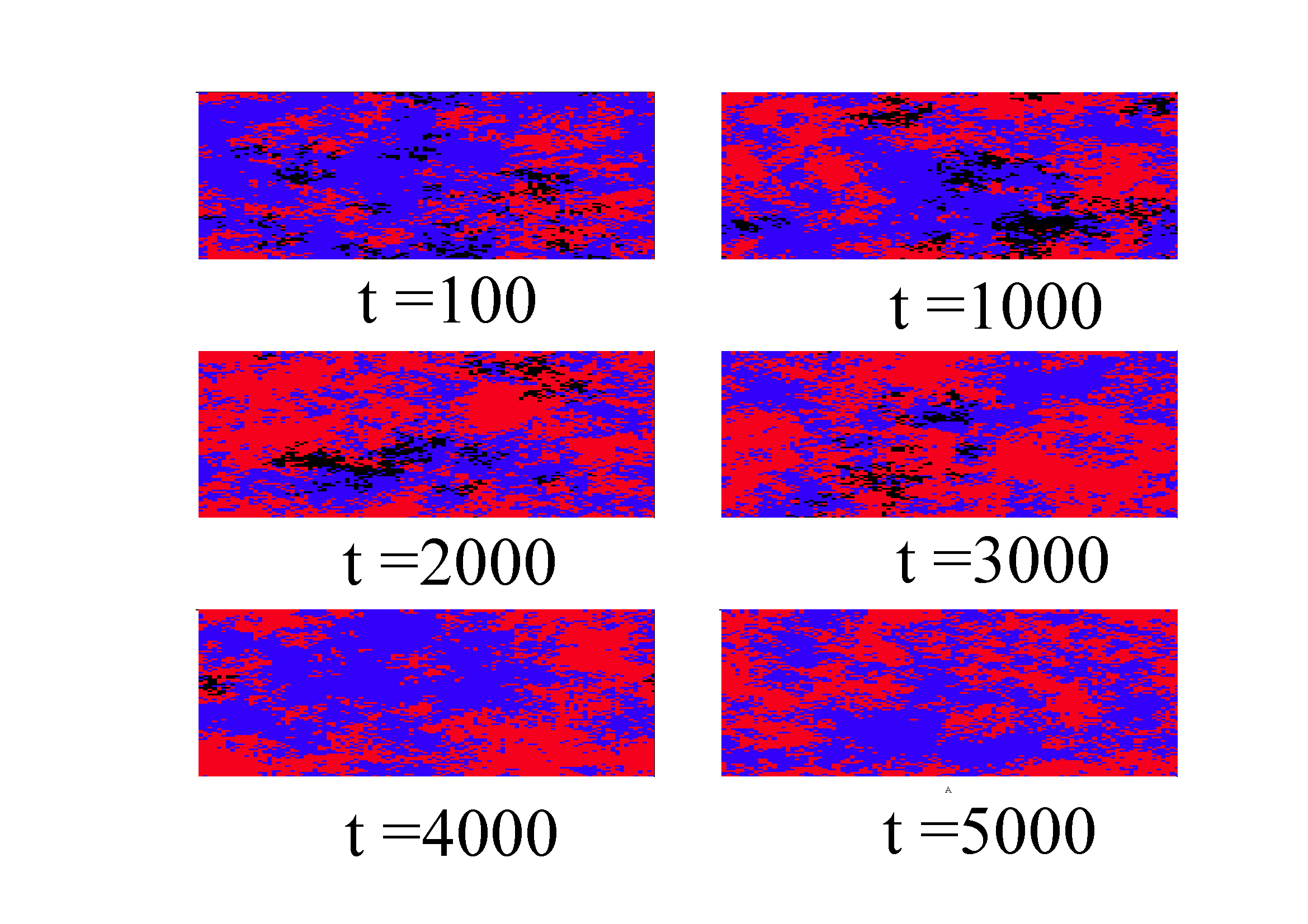}
\end{center}
\caption{Snapshots of spatial distribution of strategies over time: black:
greedy, red: moderate, and blue: conservative. }
\label{Fig:Snapshots}
\end{figure}

We have monitored the density of strategies as function of time for
different mobility values $\alpha$. First we consider the density of
strategies for one only run as shown in Fig. \ref{Fig:one_run}. The results
suggest that for low mobility, only conservative players survive. However as 
$\alpha$ increases the moderate players start increasing and the system pass
by a coexistence between conservative and moderate players, which is
followed by higher extension of survival of the greedy players that are
always extinct. For $\alpha >\alpha _{c}$, with $0.2<\alpha _{c}<0.3$ there
is a dominance of moderate players. Looking at $\alpha =0.3$, it is
interesting to observe, considering one run only, that there is a region of
coexistence between conservative and moderate players. It is interesting to
observe a snapshot of spatial distribution of strategies which illustrate
this coexistence as described in Fig. \ref{Fig:Snapshots}.

\begin{figure}[tbh]
\begin{center}
\includegraphics[width=1.0%
\columnwidth]{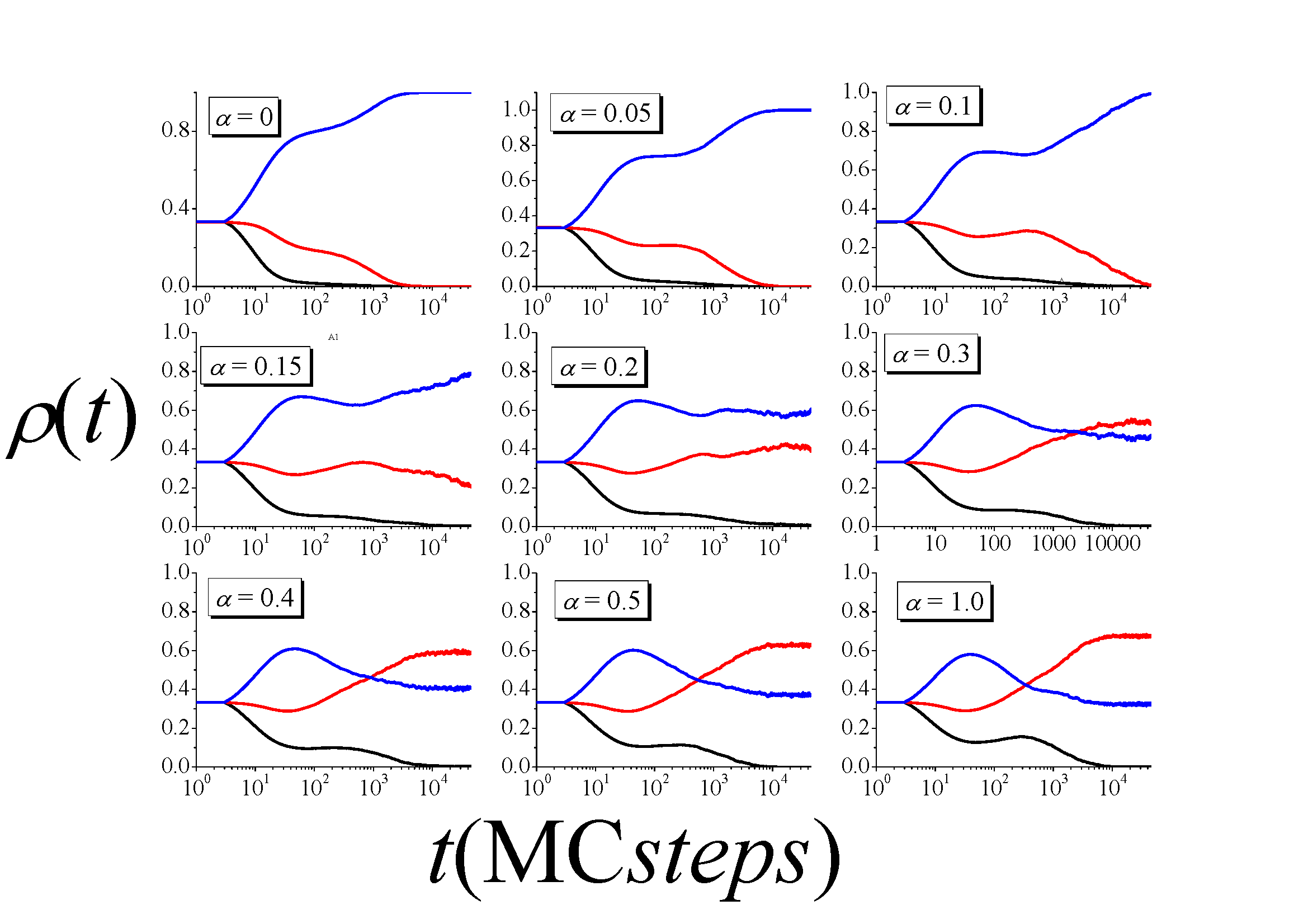}
\end{center}
\caption{Density of strategies of the MSSUG averaged over 300 runs.}
\label{Fig:300_runs}
\end{figure}

It is interesting to run again the experiment on the density of strategies
considering many different runs. So, we consider the density average over $%
N_{run}=300$ different runs. The results are shown in Fig. \ref{Fig:300_runs}%
.

\begin{figure}[tbh]
\begin{center}
\includegraphics[width=1.0\columnwidth]{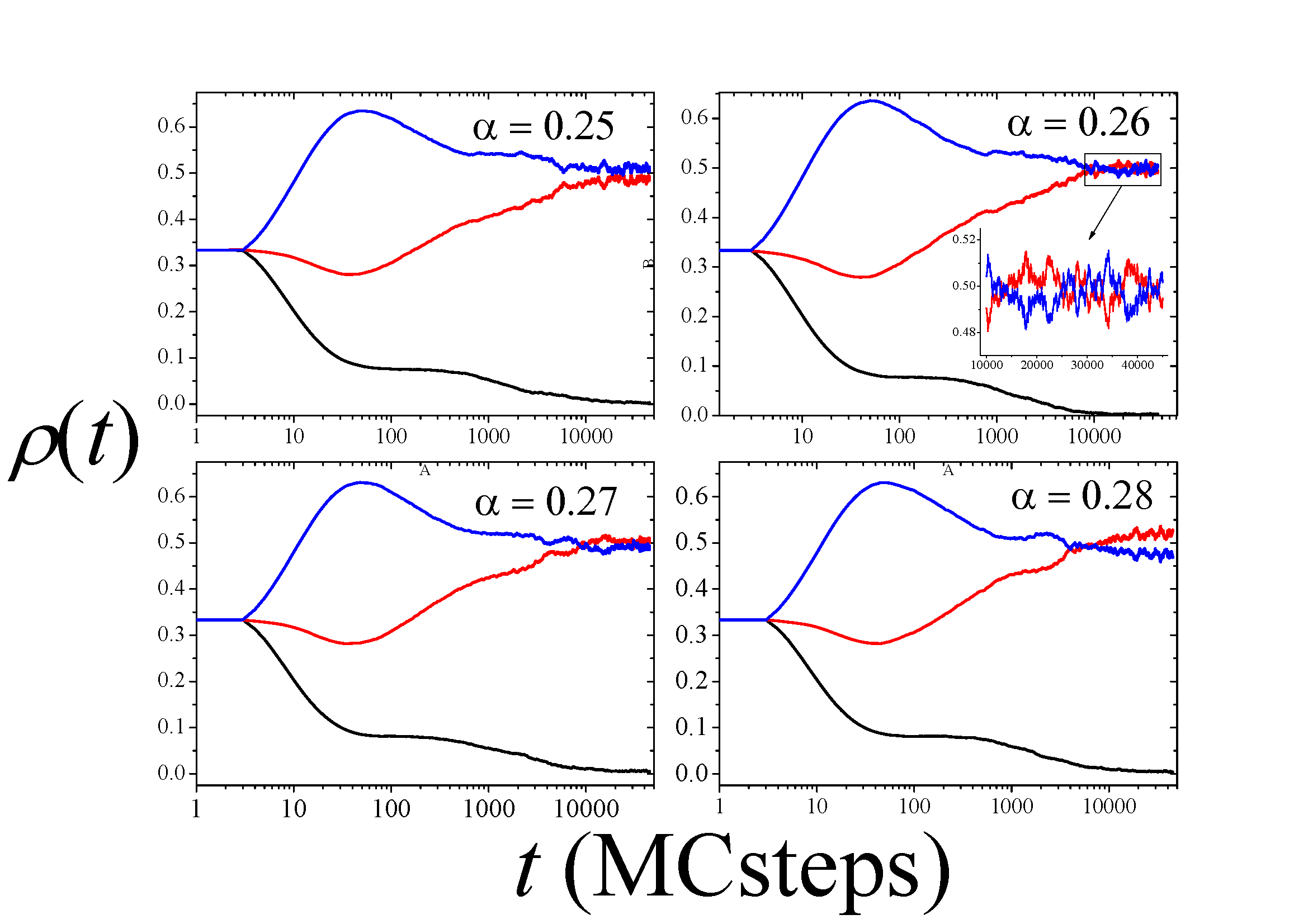}
\end{center}
\caption{Refinement of evolution of density of strategies. The inversion of
dominance occurs for $\protect\alpha \approx 0.26$, where we clearly has a
coexistence between the conservative and moderate strategies. }
\label{Fig:alfa_critico}
\end{figure}

We can observe in more detail the inversion of dominance between
conservative and moderate agents is indeed occurring the interval $[0.2,0.3]$%
, but it is interesting to refine the interval. Thus, we carried out
simulations with $\Delta \alpha =0.01$, between $\alpha =0.2$ and $0.3$, and
the inversion of dominance occurs exactly at $\alpha _{c}\approx 0.26$ as
can be observed in Fig. \ref{Fig:alfa_critico}. At this value we can see
clearly a coexistence between the two strategies. A highlight of coexistence
region is shown for $\alpha =0.26$.

If the mobility affects the dominant strategy of the agents, where the
conservative agents dominate the moderate agents, for low mobility, while
for high mobility the moderate strategy is the winner, this inverting the
game, it is interesting to explore other parameters.

We thus choose four parameters:

\begin{enumerate}
\item \textbf{Average payoff of proposers} 
\begin{equation*}
\left\langle P_{pr}(t)\right\rangle =\frac{1}{N\cdot N_{run}}%
\sum_{i=1}^{N}\sum_{j=1}^{N_{run}}P_{pr}(i,j;t)
\end{equation*}%
where $P_{pr}(i,j;t)$ corresponds to payoff of player $i$, only gaining as
proposer, for sample $j$ in the simulation, at time $t.$

\item \textbf{Average payoff of responders}, which is similarly calculated
as 
\begin{equation*}
\left\langle P_{ac}(t)\right\rangle =\frac{1}{N\cdot N_{run}}%
\sum_{i=1}^{N}\sum_{j=1}^{N_{run}}P_{ac}(i,j;t)
\end{equation*}

where $P_{ac}(i,j;t)$ corresponds to payoff of player $i$, only gaining as
responder, for sample $j$ in the simulation, at time $t$.
\end{enumerate}

By defining 
\begin{equation*}
P(i,j;t)=P_{pr}(i,j;t)+P_{ac}(i,j;t)
\end{equation*}%
the total payoff of player $i$, for sample $j$ at time $t$ we have propose
our third parameter:

\begin{enumerate}
\item[3.] \textbf{Gini coefficient of total payoff}, which is calculated as%
\begin{equation*}
G(t)=\frac{1}{N_{run}}\sum_{j=1}^{N_{run}}G(j;t)
\end{equation*}%
where $0\leq G(j;t)\leq 1$ is the Gini coefficient \cite{Gini1921} of the
total payoff of the agents for sample $j$, which was already used by
ourselves in the context of game theoretical models (see for example \cite%
{Silva2016,SilvaJTB}), 
\begin{equation*}
G(j;t)=\frac{1}{(N-1)}\left[ N+1-2\left( \frac{\sum_{i=1}^{N}(N+1-i)P(i,j;t)%
}{\sum_{i=1}^{N}P(i,j;t)}\right) \right] 
\end{equation*}%
only to measure statistical heterogeneity of payoffs and no social context
must be here explored, exactly as variance or other index, as for example
the Pietra index \cite{Eliazar2010}. And finally:

\item[4.] \textbf{Average offer}, which is calculated as 
\begin{equation*}
\left\langle O(t)\right\rangle =\frac{1}{NN_{run}}\sum_{i=1}^{N}%
\sum_{j=1}^{N_{run}}O(i,j;t)
\end{equation*}

where $O(i,j;t)$ is the offer of player $i$, for sample $j$, at time $t$.
\end{enumerate}

And the question here is, what happens with these four parameters as a
function of time? Moreover, what the is diffusion effect on them?

\begin{figure}[tbh]
\begin{center}
\includegraphics[width=1.0\columnwidth]{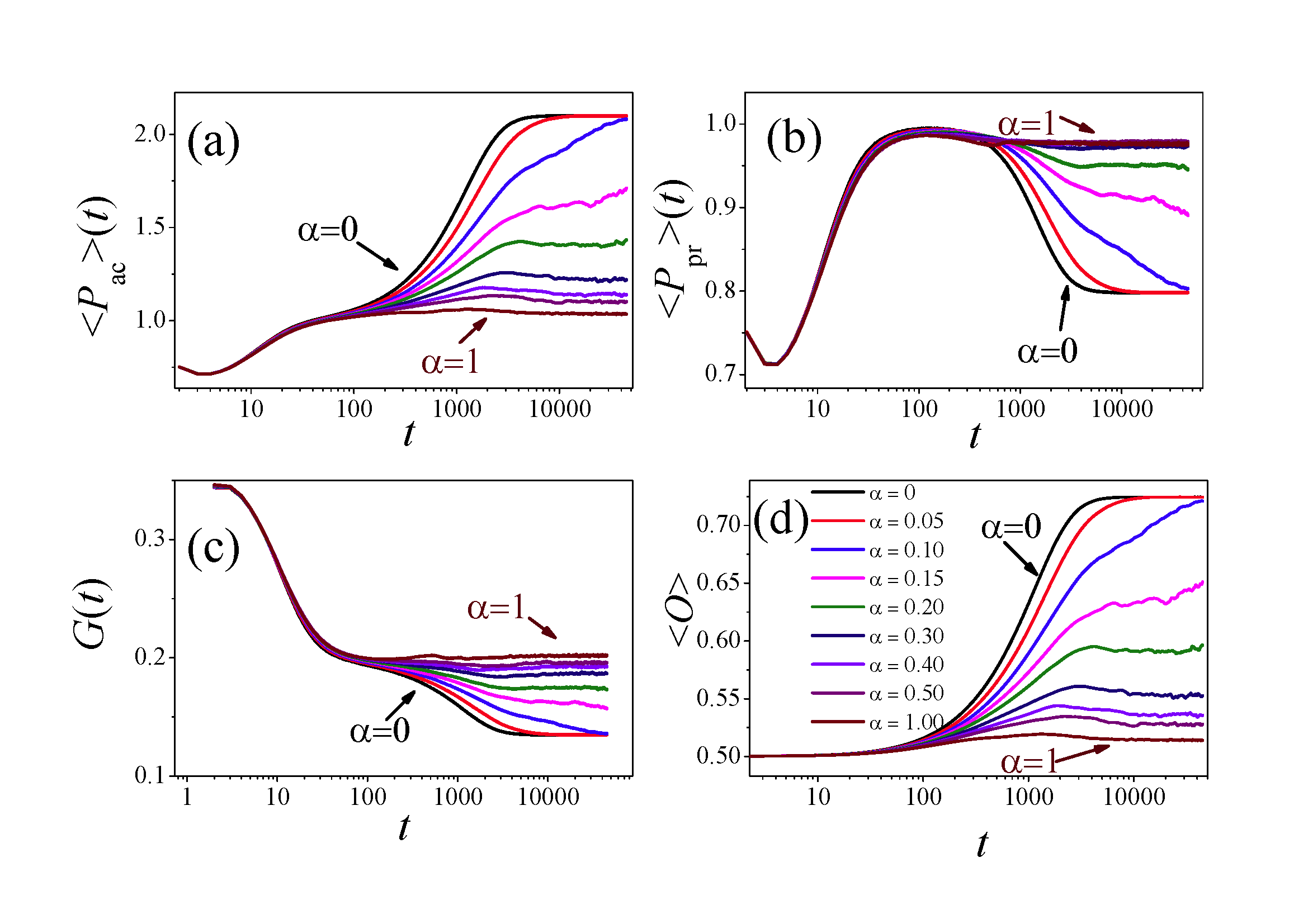}
\end{center}
\caption{Temporal evolution of four parameters that describe the
fluctuations on payoff and offer of agents under spatial diffusion effects.
(a) Average payoff of responders, (b) Average payoff of proposers (c) Gini
coefficient, and (d)\ Average offer. }
\label{Fig:parameters}
\end{figure}

As one can observe in Fig. \ref{Fig:parameters} (a) the $\left\langle
P_{ac}(t)\right\rangle $ increases as $t$ increases, but this increase is
reduced as the mobility of agents grows. Differently, the mobility is good
for the average payoff of the responders since $\left\langle
P_{pr}(t)\right\rangle $ for large times gets higher, the higher $\alpha$
is, Fig. \ref{Fig:parameters} (b). We know that low mobilities has a
dominance of conservative players, while high mobilities shows the dominance
of moderate ones. This is reflected by average offers, Fig. \ref%
{Fig:parameters} (d), since conservative players require more levels of
acceptance, the offer is increased to result in more acceptances, in this
case there is a fixed point $O_{\infty }\approx 0.724$ for the average
offer. However, as mobility increases, the average offer reaches fairer
steady states. Finally, we analyse the Gini coefficient that decreases as
function of time Fig. \ref{Fig:parameters} (c). Mobility makes the
heterogeneity of the payoff larger. The transition between the dominances is
again observed, since the Gini coefficient has a fixed point ($G_{\infty
}=0.135$) reached only for low mobilities (below $\alpha _{c}=0.26$). Thus
the parameters show non-trivial behaviour of the game which corroborate the
transition previously observed with density of strategies.

\section{Conclusions}

\label{Section:Conclusions}

In this paper, we considered a reactive version of ultimatum game where the
players regulate their offers and can assume three different strategies:
greedy ($G$), moderate ($M$), and conservative ($C$) that differ on the
level of greed. First, we show how to compose portfolios (probabilities of
acting by choosing among these three strategies) that lead to fairer
situations. Analytical results and a diagram representing this fairness is
presented. Further, by including Darwinian and spacial aspects, we studied a
multi-agent system of the model which shows that mobility has an important
role over the dominance of strategies. We show that there exist a transition
between dominances on the diffusion probability $\alpha $, for $\alpha <0.26$%
, $C\succ M$, while for $M\succ C$ for $\alpha >0.26$. A coexistence between
these two strategies can be observed exactly in $\alpha =0.26$. This
transition is reflected in four different parameters related to the offer,
payoff, and the heterogeneity among the agents.

\textbf{Acknowledgements --} Roberto da Silva thanks CNPq for financial
support under grant numbers: 311236/2018-9,and 424052/2018-0. Luis Lamb is
partly supported by CNPq and by CAPES - Coordena\c{c}\~ao de Aperfei\c{c}%
oamento de Pessoal de N\'{\i}vel Superior - Financed Code 001. This work was
partly developed using the computational resources of Cluster Ada, IF-UFRGS.

\bigskip

\bigskip

\end{document}